\documentclass[twocolumn,showpacs,preprintnumbers,amsmath,amssymb]{revtex4}

\usepackage{graphicx}
\usepackage{dcolumn}
\usepackage{bm}

\begin{document}


\title{Investigation of the Jahn-Teller Transition in TiF$_3$ using Density Functional Theory.}

\author{Vasili Perebeinos and Tom Vogt}
\affiliation{Department of Physics, Brookhaven National Laboratory, Upton,
New York 1973-5000 }

\date{\today}

\begin{abstract}

We use first principles density functional theory to calculate
electronic and magnetic properties of TiF$_3$ using the full
potential linearized augmented plane wave method. The LDA
approximation predicts a fully saturated ferromagnetic metal and
finds  degenerate energy minima for high and low symmetry
structures. The experimentally observed Jahn-Teller phase
transition at T$_c$=370K can not be driven by the electron-phonon
interaction alone, which is usually described accurately by LDA.
Electron correlations beyond LDA are essential to lift the
degeneracy of the singly occupied Ti t$_{2g}$ orbital. Although
the on-site Coulomb correlations are important, the direction of
the t2g-level splitting is determined by the dipole-dipole
interactions. The LDA+U functional predicts an aniferromagnetic
insulator with an orbitally ordered ground state. The input
parameters U=8.1 eV and J=0.9 eV for the Ti 3d orbital were found
by varying the total charge on the TiF$_6^{2-}$ ion using the
molecular NRLMOL code. We estimate the Heisenberg exchange
constant for spin-1/2 on a cubic lattice to be approximately 24 K.
The symmetry lowering energy in LDA+U is about 900 K per TiF$_3$
formula unit.
\end{abstract}

\maketitle

\section{\label{sec1}Introduction}

There is an ongoing interest in phase transitions in perovskite
based materials. Above ${\rm T_c}\approx$370 K the trifluoride
TiF$_3$ has the cubic framework perovskite structure AMX$_3$, with
no A cations present. Each Ti is at the center of a corner sharing
fluorine octahedra MX$_6$. At low temperatures the structure
becomes rhombohedral. This symmetry lowering can to a first
approximation be characterized by a titling of the rigid MX$_6$
octahedra about the threefold axis. The nominal valance of the
titanium ion is $3+$, with one 3d electron occupying the triply
degenerate $t_{2g}$ orbital in the high temperature phase. One may
expect a Jahn-Teller instability in TiF$_3$. Indeed  in the
distorted structure titanium has a $D_{3d}$ local environment, in
which the $t_{2g}$ orbitals are split into an $a_{1g}$ and a
doublet $e_{g}$ orbital. The trigonal distortions have been
intensively discussed in the context of other $t_{2g}$ compounds,
like FeO \cite{Isaak} and LaTiO$_3$ \cite{Mochizuki}.

We use Density Functional Theory (DFT) to investigate the cubic to
rhombohedral structural phase transition which has been
established by X-ray diffraction \cite{Mogus,Kennedy}. Many
trifluoride MF$_3$ compounds (M=Al, Cr, Fe, Ga, In, Ti, V) exhibit
this structural phase transition
\cite{Mogus,Kennedy,Ravez,Daniel}. It is believed that the
observed transition in TiF$_3$ is of a ferroelastic nature
\cite{Mogus}. In most of those materials except for TiF$_3$ and
MnF$_3$ there are no partly filled $d$ shells. The driving
mechanism for the ferroelastic transitions can be pictured as the
formation of the dipole moments on fluorine atoms due to the
asymmetric distribution of the $2p$ electron density in the
distorted structure. The long range dipole-dipole interaction
favors a distorted structure with anti-polar arrangement of
dipoles \cite{Allen2}. The Local Density Approximation (LDA)
captures both the long-range and the short interactions on the
same footing and predicts a distorted structure to have a minimum
energy in AlF$_3$ \cite{Chen}.

In the present studies we choose the TiF$_3$ system with a partly
filed $t_{2g}$ shell. The Jahn-Teller energy lowering due to the
lifting of the $t_{2g}$ orbital degeneracy in the distorted
structure should add up to the long range dipole formation energy.
Unexpectedly, we find the total energies of the high and low
temperature phases to be identical within the errors of
calculations. The failure of LDA to explain this phase transition
is due to the fact that transition metal $d$ electrons are not
adequately described by LDA. Although the dipole-dipole long range
interaction favors the distorted structure, however the electron
kinetic energy of the $t_{2g}$ one third filled band is increased
due the diminishing of the hopping integral ($t$) between the
neighboring Ti atoms due to the octahedra tilting. The two effects
mutually cancel each other resulting in degenerate minimums of the
potential energy surface.

It was realized some time ago that LDA tends to underestimate the
Coulomb repulsion of electrons occupying different orbitals of the
same $d$-shell \cite{Anisimov}. In particular, this leads to equal
occupations of different orbitals of the same manifold and
prevents the stabilization of the orbitally ordered solutions. The
LDA+U functional, however, generates an orbital dependent
potential which favors solutions with broken orbital degeneracy.
LDA+U calculations indicate that the low temperature phase has a
by about 900K lower energy per TiF$_3$ than the high temperature
phase. One of the roles of the Coulomb interaction is to
suppresses the electron kinetic energy of the partly filled
$t_{2g}$ bands. The bandwidth narrowing in the distorted structure
effectively reduces the restoring force to rotate the octahedra
which facilitate stabilization of the distorted structure.
The electronic ground state is orbitally ordered in the low
temperature phase with one electron occupying the a$_g$ orbital
oriented along the rhombohedral direction forming a Heisenberg
spin 1/2 lattice coupled antiferromagnetically to its neighbors
{\it via} a superexchange mechanism.


In this work we use the molecular NRLMOL code \cite{Mark} to
calculate $U$ by varying the occupation number of the Ti $d$
orbital of the TiF$_6^{2-}$ ion. These calculations are
significantly less computationally demanding compared to the
supercell approaches \cite{Anisimov,Solovyev2,Pickett}and since
only the first neighboring fluorines mostly contribute to the
screening a good estimate for parameters $U$ and $J$ can be
readily achieved in the calculations.

The Hund interaction and the electron kinetic energy are
realistically described with LDA. The TiF$_3$ compound is
predicted to be a ferromagnetic metal with a fully saturated
magnetic moment of 1 $\mu_B$ per formula unit. The exchange Hund
energy is large and the Stoner criterium is satisfied. The
electron-phonon interaction lifts the on site degeneracy of the
$t_{2g}$ orbitals and for a sufficiently strong coupling a gap in
the spectrum can open. We estimate the electron-phonon coupling
($\approx$ 2 meV/degree) to be insignificant to open the gap and
to drive the phase transition in TiF$_3$. Although the on-site
Coulomb correlations are important, the direction of the t2g-level
splitting is determined by the long range dipole-dipole
interaction \cite{Allen2}. Many successes and failures of LSDA in
transition metal oxide compounds have been summarized, for
example, in the review article \cite{Solovyev1}.

\section{\label{sec2}Method of calculation}

The purpose of this paper is to elucidate the electronic structure
of TiF$_3$ and explain the structural phase transition using the
DFT method. This is done using the full potential linearized
augmented plane wave (FP-LAPW) method \cite{Singh1} with local
orbital extensions \cite{Singh2} in the WIEN2k implementation
\cite{Blaha}. The LDA Perdew Wang \cite{Perdew}
exchange-correlation potential was used. Well converged basis sets
and Brillouin zone sampling were employed. The crystal structure
was reported by Kennedy and Vogt \cite{Kennedy}. It is cubic at
high temperature with a unit cell volume 58.8 (\AA$^3$). A
rhombohedrally distorted structure of the space group $R$-3c can
be characterized by three parameters: (1) volume 56.5 (\AA$^3$),
(2) the octahedra tilt $\phi=13^0$, and (3) a $c/a=1.0315$ ratio,
which measures the distortion along the rhombohedral direction.
There are two titanium and 6 fluorine atoms in the rhombohedral
unit cell. The fluorine ions are sitting in the $6e$ sites $((x,
-x+1/2, 1/4),{\rm etc.})$, and $\delta=x-0.75$ is the deviation
from cubic positions. The octahedra tilting angle  is related to
$\delta$ by $\tan(\phi=2\sqrt2\delta$. The calculations for {\it
both} the high and low symmetry phases were performed using two
formula per unit cell and the same k-point mesh, with 292 special
k-points in the irreducible Brillouin zone. A tetrahedron method
was used for integration over the Brillouin zone.

\section{\label{sec3}LDA results}

\begin{figure}
\includegraphics[height=2.62in,width=2.37in,angle=0]{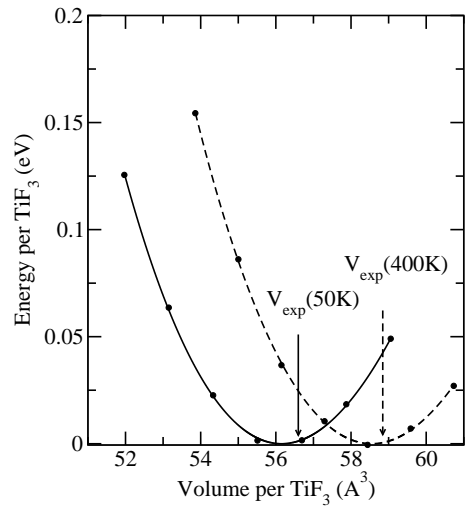}
\caption{\label{fig1}Calculated total energy of ferromagnetic
high-T phase ($\phi=0$, $c/a=1$ - dashed line) and low-T phase
($\phi=10^0$, $c/a=1.0315$ - solid line) as a function of volume.}
\end{figure}

We first calculate the cubic phase by fixing $\phi=0$ and $c/a=1$
in the rhombohedral structure of the space group $R$-3c. LDA finds
a ferromagnetic ground state solution with a fully saturated
magnetic moment of 1 $\mu_B$ per Ti. The volume optimization is
shown in Fig. \ref{fig1}. The equilibrium volume $V_0=58.6$
\AA$^3$ is in excellent agreement with the experiment
\cite{Kennedy}. The bulk modulus of 111 GPa is extracted by
fitting a
 Murnaghan \cite{Murnaghan} equation of state.
The $c/a$ ratio was fixed to the experimental value 1.0315 at T=10
K and the two parameters are relaxed to find the minimum of the
low symmetry structure. The energy minimization with respect to
the volume for the fixed $c/a$=1.0315 and $\phi=10^0$ is shown on
Fig. \ref{fig1} and yields
 $V_m=56.2$ \AA$^3$.

\begin{figure}
\includegraphics[height=2.608in,width=2.43in,angle=0]{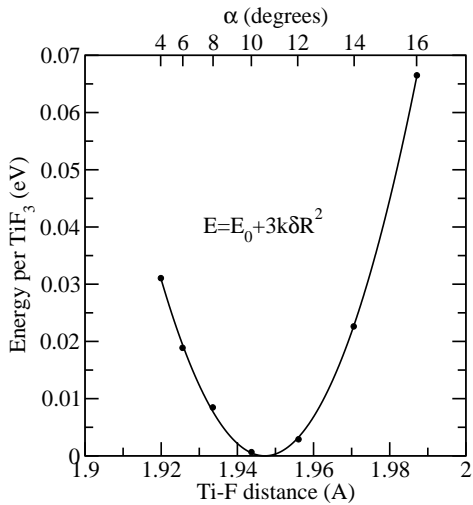}
\caption{\label{fig2} Calculated total energy of ferromagnetic
low-T phase as a function of the tilt angle and fixed volume
($V_m$) and c/a=1.0315 ratio. The optimal Ti-F distance is 1.947
\AA \ and the spring constant is 13.9 eV/\AA$^2$.}
\end{figure}

Further optimization with respect to the tilt angle for a fixed
minimum volume $V_m$ and the same $c/a$ ratio did not
significantly alter the ground state energy. The optimal tilt
angle is $\phi_0=10.5^0$. The Ti-F bond is the most rigid bond in
the structure, and therefore it is convenient to plot the energy
versus the Ti-F bondlength. A parabola fit to the energy variation
shown on Fig.\ref{fig2} is excellent and yields a spring constant
of the single bond $K=13.9$ eV/\AA. The Ti-F bond length $d$ in
$R3c$ crystal structure is
\begin{eqnarray}
d^{2}=a_h^2(1/12+\delta^2+(c/a)^2/24)
\label{TiF}
\end{eqnarray}
where $a_h=a\sqrt{2}$ is the hexagonal a
lattice constant. Knowing the spring constant for the bond stretching
and equilibrium volume $V_0$ the bulk modules can be estimated as
$B=K/(6V^{1/3})\approx96$ GPa.

The total energy differences of the high- and low-T phases is beyond the
accuracy of calculations. Despite the excellent agreement with the
experimental structural parameters, the LDA total energy analysis does not
explain the high temperature of the observed structural phase transition.

\begin{figure}
\includegraphics[height=2.18in,width=2.85in,angle=0]{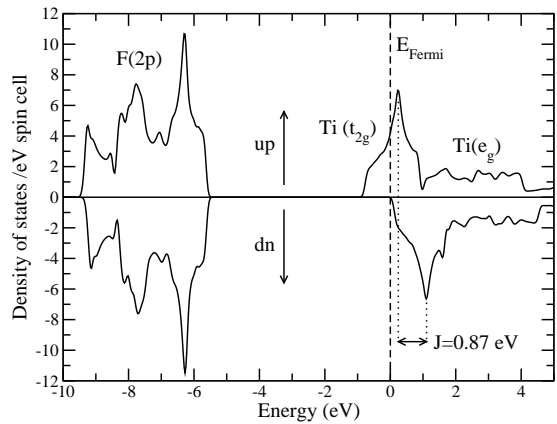}
\caption{\label{fig3} Density of states of the ferromagnetic
solution for high-T phase. The $t_{2g}$ and $e_g$ bands are split
by the crystal field energy $\approx 2$ eV. The $t_{2g}$ bandwidth
is about 2 eV, which corresponds to hopping parameter $t=0.25$ eV.
The Hund coupling $J=0.87$ eV can be estimated from the relative
position of the spin up and spin down peak positions.}
\end{figure}

\begin{figure}
\includegraphics[height=2.48in,width=2.48in,angle=0]{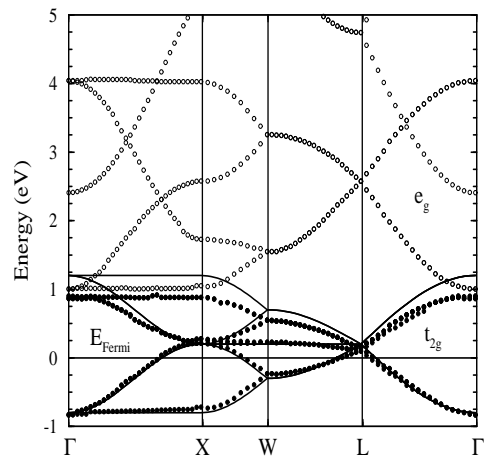}
\caption{\label{fig4}
Calculated band structure along the high symmetry line in
the high symmetry structure: $e_g$ bands open circles, $t_{2g}$ bands closed
circles. The solid line is a tight binding fit, which uses
a single Slater-Koster parameter $t=(dd\pi)=0.25$ eV. Fermi level is
at zero energy.
}
\end{figure}

The densities of states are shown on Fig. \ref{fig3}. The $t_{2g}$ band lies at
the Fermi level and it is one third filled. The Hund energy dominates the
kinetic energy such that the Stoner criterium is satisfied and the  material
becomes magnetic.
The band structure of the Ti t$_{2g}$ electrons can easily be understood
with a nearest-neighbor two-center Slater-Koster \cite{Slater} model.
The orbitals $|\alpha\beta>$ have hopping matrix elements with themselves
along the $\alpha$ and $\beta$ directions with amplitude t=(dd$\pi$).
The tight binding fit in the high temperature phase with $t=0.25$ eV is
shown on Fig. \ref{fig4}.

In the low temperature phase the Ti-F-Ti bond angle is smaller than
180$^0$, which reduces the hopping and narrows the band width.
The onsite orbital degeneracy is lifted by the octahedra tilting.
In the $D_{3d}$ local Ti site symmetry the $t_{2g}$ manifold splits
into:

\begin{eqnarray}
a_g&=&\frac{|xy>+|yz>+|zx>}{\sqrt{3}}
\nonumber\\
e_{g1}&=&\frac{|yz>-|zx>}{\sqrt{2}}
\nonumber\\
e_{g2}&=&\frac{2|xy>-|yz>-|zx>}{\sqrt{6}}
\label{orb}
\end{eqnarray}

The bandwidth in the tilted structure $\phi=\phi_0$ corresponds to
reduced t=0.225 eV and the electron-phonon Jahn-Teller energy gap
at $\Gamma$ point is 6 meV. From the simple tight binding model
with these parameters the kinetic energy loss due to the band
narrowing effect is 18 meV and it is not even compensated by the
lifting of the orbital degeneracy. The total LDA energy includes
the lattice and electronic degrees of freedom and predicts
degenerate minimums within the errors of calculations (Fig.
\ref{fig1}). The LDA failure to predict the cubic-to-rhombohedral
phase transition can be rationalized by the lack of electron
correlation effects in the LDA method.

\section{Parameter U (J) calculation for LDA+U Method}

In extending the LDA method to account for correlations resulting
from on-site interactions Anisimov, Zaanen, and Andersen (AZA)
\cite{Anisimov} chose to refine LDA by including an
orbital-dependent one-electron potential to account explicitly for
the important Coulomb repulsions not treated adequately in the LDA
approach. This was accomplished in accordance with Hartree-Fock
theory by correcting the mean-field contribution of the $d-d$
on-site interaction with an intra-atomic correction. This
correction has been applied in slightly different ways. We use the
SIC LDA+U functional \cite{SIC} as implemented in the WIEN2k
package.


In this work we use a single TiF$_6^{2-}$ ion to calculate
parameters $U$ and $J$.  In these calculations the screening due
to the nearest neighboring fluorines is taken into account and
there are no other Ti atoms for the $d$ electron to hop to. The
TiF$_6^{2-}$ ion forms an octahedra with a Ti-F bond of 1.93 \AA \
\ with the Ti atom placed in the center. The total energy and $d$
orbital chemical potential of the TiF$_6$ can be modelled by the
single site Hubbard model:
\begin{eqnarray}
E&=&E_0-\varepsilon(n_{\uparrow}+n_{\downarrow})+
\frac{U}{2}(n_{\uparrow}+n_{\downarrow})^2-
\frac{J}{2}(n_{\uparrow}^2+n_{\downarrow}^2)
\nonumber\\
\mu_{\uparrow}&=&\frac{\partial E}{\partial n_{\uparrow}}=
-\varepsilon+U(n_{\uparrow}+n_{\downarrow})-Jn_{\uparrow}
\label{ion}
\end{eqnarray}
where $n_{\uparrow}$ and $n_{\downarrow}$ are occupation numbers
of the triply degenerate $t_{2g}$ molecular orbital. We use two
sets of self-consistent calculations to determine the parameters
$U$ and $J$.
The quadratic total energy and linear chemical potential fits to
the nonmagnetic calculations give $U_1^{eff}=U-J/2=7.66$ eV. From
the fully polarized magnetic calculation the quadratic energy and
linear $\mu_{1/2}$ fit we find $U_2^{eff}=U-J=7.24$ eV, such that
$U=8.08$ eV and $J=0.84$ eV. The Hund coupling parameter $J$
should be compared with the spin up and down energy bands
splitting in ferromagnetic LDA calculations Fig. \ref{fig3}, which
is $J=0.87$ eV.

\section{LDA+U Results}
\begin{figure}
\includegraphics[height=2.608in,width=2.26in,angle=0]{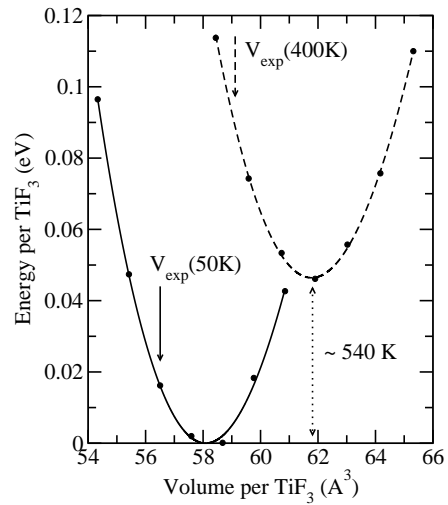}
\caption{\label{fig6} LDA+U calculations of the antiferromagnetic
high-T phase ($\phi=0^0$, $c/a=1$ - dashed line) and low-T phase
($\phi=12^0$, $c/a=1.0315$ - solid line) as a function of the
volume of TiF$_3$. Coulomb and exchange parameters were chosen
U=8.1 eV and J=0.9 eV. The optimal volumes of 61.8 \AA$^3$ and
58.1 \AA$^3$ agrees well with experimental results indicated by
arrows.}
\end{figure}

The volume optimization of the high symmetry phase gives
$V_0=61.8$ \AA$^3$ using two formula per unit cell with $R-3c$
point group symmetry operations and fixed $\phi=0^0$ and $c/a=1$.
The bulk modulus extracted from the Murnaghan fit shown on
Fig.\ref{fig6} is 110 GPa. The electron correlations on the Ti $d$
orbitals reduces the bandwidth ($4t/U$), such that the kinetic
energy variation with respect to the lattice constant is smaller
in the LDA+U functional resulting in a larger equilibrium lattice
constant compared to the LDA result.

\begin{figure}
\includegraphics[height=2.608in,width=2.42in,angle=0]{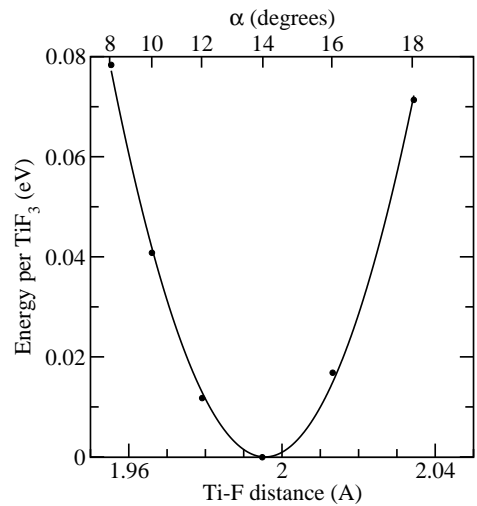}
\caption{\label{fig7} LDA+U calculations for the antiferromagnetic
low-T phase as a function of the tilt angle and fixed volume $V_m$
and c/a=1.0315 ratio. The optimal angle is 14.1$^0$, which
corresponds to Ti-F distance 1.996 \AA. The spring constant is
15.95 eV/\AA$^2$.}
\end{figure}

The low symmetry optimization was first done
with respect to volume for a fixed experimental tilt angle $\phi=12^0$
and the ratio $c/a=1.0316$. The optimal volume is 58.1 \AA$^3$.
Further optimization with respect to the
tilt angle for the fixed equilibrium
volume gives an additional energy gain of 130 K per Ti such that the
distorted structure with the optimal tilt angle of 14$^0$ is lower in energy by
675 K per Ti.  The parabola fit to the Ti-F bond length variation shown
on Fig. \ref{fig7} gives spring constant $K=15.95$ eV/\AA$^2$,
The corresponding bulk modulus $B=K/(6V^{1/3})\approx108$ GPa is an
in excellent agreement with Murnaghan fit Fig. \ref{fig6}.

\begin{figure}
\includegraphics[height=2.608in,width=2.50in,angle=0]{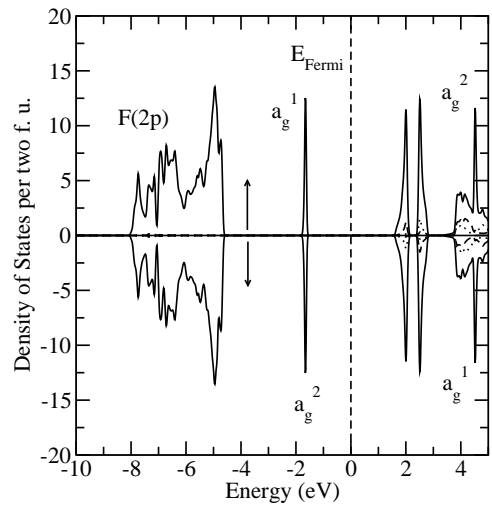}
\caption{\label{fig8}
Density of States per two formula units per eV. The two electrons in the
unit cell fill the $a_g^1$ spin up orbital on atom one and
the $a_g^2$ spin down orbital on the second atom peaked at -1.64 eV.
The two lowest unoccupied peaks (at 2.01 eV and 2.52 eV) are due to the  $e_g$ orbitals
of the $t_{2g}$ manifold. The lower energy orbital is due to the
$e_g$ orbital of the same atom and spin as the occupied $a_g$ orbital.
The broad bands 2 eV higher in energy shown by the dashed and dotted lines
are due to the $e_g=\{x^2-y^2,3z^2-r^2\}$
orbitals split by crystal field.
The empty $a_g$ orbital with opposite spin at 4.53 eV forms a resonance in the
crystal field $e_g$ background.}
\end{figure}

The density of states for the distorted structure is shown on Fig.
\ref{fig8}. The fluorine $2p$ states are not much affected by the
correlations. Whereas dramatic differences are seen for the Ti $d$
states. First, LDA+U  predicts a gap for {\it both} high-T and
low-T phases in the $t_{2g}$ band, which is split into three
distinct narrow peaks originated from the $a_g$ and two $e_g$
orbitals. The imposed rhombohedral symmetry of the supercell makes
the lowest energy $a_g$ orbital pointing along the $(111)$
direction to be fully occupied with a spin up electron on one atom
and a spin down on the other atom coupled antiferromagnetically.
The electronic properties of TiF$_3$ can be described by the
Hubbard Hamiltonian:
\begin{eqnarray}
{\cal H}_{\rm el}&=&\sum_{i,\alpha\beta,\sigma}
-t\left(c^{\dagger}_{\alpha\beta,\sigma,i}c_{\alpha\beta,\sigma,i+\alpha}
+c^{\dagger}_{\alpha\beta,\sigma,i}c_{\alpha\beta,\sigma,i+\beta}\right)
\nonumber\\
&+&\sum_{i,\alpha\beta,\sigma}
\frac{U}{2}
\left(n_{\alpha\beta,\sigma,i}n_{\alpha\beta,-\sigma,i}\right)
\nonumber\\
&+&\sum_{i,\alpha\beta\ne\alpha'\beta',\sigma,\sigma'}
\frac{U'-J\delta_{\sigma,\sigma'}}{2}
\left(n_{\alpha\beta,\sigma,i}n_{\alpha\beta,\sigma',i}\right)
\label{hmodel}
\end{eqnarray}
The first term is a kinetic energy term described in Section
\ref{sec3}. To place two electrons with opposite spins costs
energy $U$, if they occupy the same orbital, or $U'$, if they are
on different orbitals. If the spins of two electrons on different
orbitals are aligned the energy cost is reduced by $J$. In the
atomic limit the relation $U=U'+2J$ holds. Then the splitting
between the $a_g$ and $e_g$ orbitals for a single site Hubburd
model would be $U'-J\approx 5.6$ eV. The atomic limit value of
splitting is larger from the complete $LDA+U$ solution $\approx 4$
eV (see Fig. \ref{fig8}) where only a fraction of $d$ electrons
inside the muffin-tin sphere  experiences the LDA+U potential. The
second order perturbation theory of the Hamiltonian (Eq.
\ref{hmodel}) with respect to $t/U$ predicts an insulating
antiferromagnetic ground state solution for parameters $U=8.1$ eV,
$J=0.9$ eV and $t=0.22$ eV.

\vspace{15pt}

\begin{figure}
\includegraphics[height=2.608in,width=2.63in,angle=0]{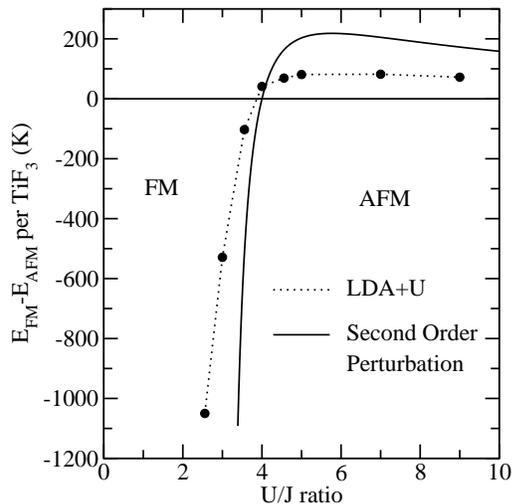}
\caption{\label{fig9}
Energy difference of the total energy of ferromagnetic and
antiferromagnetic solutions as a function of U/J ratio. The
solid line is a second order perturbation theory for the fixed values
J=0.9 eV and t=0.225 eV. The dots are $LDA+U$ results with fixed $J=0.9$ eV
and optimal low temperature geometry.}
\end{figure}

The superexchange energies per Ti of the antiferromagnetic Neel state and
ferromagnetic solutions,
with $a_g$ orbital Eq. \ref{orb} occupied in both cases, are:
\begin{eqnarray}
E_{\rm AFM}&=&-\frac{4t^2}{3}\frac{1}{U'}-\frac{8t^2}{3}\frac{1}{U}
\nonumber\\
E_{\rm FM}&=&-\frac{4t^2}{3}\frac{1}{U'-J}
\label{mfs}
\end{eqnarray}
where the first term is due to virtual hopping
of the localized electron on the
six neighboring $e_g$ orbitals. The second term in $E_{\rm AFM}$ is due to
virtual hopping of the $a_g$ orbital to the 6 neighbors with the hopping
amplitude $2t/3$, and it is missing in the ferromagnetic ground state
due to the Pauli principle. The $LDA+U$ energy differences between the
ferromagnetic and antiferromagnetic ground states and the fixed optimal
distorted geometry are shown on Fig. \ref{fig9}. Both the perturbation theory
and the $LDA+U$ results predict the antiferromagnetic to ferromagnetic
transition at $U/J\approx4$. For the large $U/J$ values $LDA+U$ underestimates
the exchange energy by a factor of $2$ and for the realistic parameter of
$U=8.1$ eV it predicts an energy difference of $72$ K between the two phases.
Neglecting the zero-point energy of the spin waves the energy difference
between the parallel and anti-parallel classical spins on a cubic lattice
is $6JS(S+1)$ \cite{Anderson}.
For spin $S=1/2$ this yields $J=16$ K and the corresponding Neel temperature
in the mean field approximation is $T_{\rm N}=1.5J=24$ K. The quantum
fluctuations reduces the Neel temperature to $T_{\rm N}=0.946J=15$ K
\cite{Sandvik}.
However, the neutron diffraction experiment \cite{Tom} did not reveal any
long range magnetic order down to 10 K.
Whether the absence of magnetic long range order in TiF$_3$ is an intrinsic
effect due to the coupling of the orbital and spin degrees of freedom or
extrinsic due to the presence of the multidomain structure observed in the
low symmetry phase by Mogus-Milankovic et al.
\cite{Mogus} requires further theoretical and experimental investigations.

%
%

\section{Pressure Anomaly}

\begin{figure}
\includegraphics[height=2.134in,width=2.70in,angle=0]{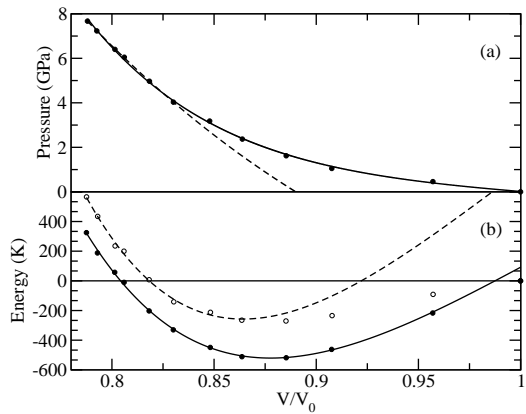}
\caption{\label{fig11}
(a) experimental $P-V$ diagram fitted by the Birch-Murnaghan equation
of state to all data points (solid line) and the first six high pressure
points (dashed line). (b) bulk modulus calculations using the
cell parameters given in table \protect{\ref{tab1}}.
Energies of LDA (LDA+U) calculations are shown by open (filled) circles.
The Birch-Murnaghan fit to the high pressure points is shown by the
dashed (solid) line for LDA (LDA+U) calculations.
The zero energy is chosen at the ambient pressure value.}
\end{figure}

X-ray diffraction measurements under pressure by
Sowa and Ahsbahs \cite{Sowa} allow us to determine the experimental
bulk modulus of TiF$_3$.
Fig. \ref{fig11} (a) shows data points along with Birch-Murnaghan
equation of state fit.
The fit to the full range of pressures predicts a physically unreasonable
small bulk modulus $B=7.5$ GPa and unphysically large coefficient
$K_2=142$ GPa$^{-1}$. The first derivative is fixed to $K_1=4$ in the
Birch-Murnaghan equation of state:
\begin{eqnarray}
P(V)=3Bf(1+2f)^{5/2}
\left(1+\frac{3f^2}{2}\left(BK_2+\frac{35}{9}\right)\right)
\label{PV}
\end{eqnarray}
where $f=0.5((V_0/V)^{2/3}-1)$ and $V_0$ is the volume at
ambient pressure.
The high pressure points (above 4 GPa) can be fitted by
Eq. (\ref{PV}) to give $B=51$ GPa and a normal $K_2=-0.005$ GPa$^{-1}$.
However, the volume at normal pressure has to be fixed to $V_0=51.9$ \AA$^{3}$,
which is 11\% smaller than the observed one.

In the present work we calculate the bulk modulus of the lower
symmetry phase using the experimental $c/a$ and $V/V_0$ ratios
\cite{Sowa}. The tilt angle was chosen to keep the Ti-F bondlength
Eq. (\ref{TiF}) constant for all pressure values. We fix
$d_{Ti-F}=1.947$ \AA\ \ and $d_{Ti-F}=1.996$ \AA\ \ for LDA and
LDA+U calculations respectively. For LDA+U calculations $V_0$ was
fixed to the experimental value \cite{Sowa} $V_0=$58.3 \AA$^3$,
while for LDA it is reduced to match the optimal volume $V_0=$56.2
\AA$^3$ to eliminate the strain effects. The results of
calculations for the unit cell parameters given in table
\ref{tab1} are shown on Fig. \ref{fig11} (b). The high pressure
points can be fitted by Eq. (\ref{PV}) to give a bulk modulus of
$B=42$ GPa for LDA and $B=29$ GPa for LDA+U functionals with
corresponding coefficients $K_2=-0.011$ GPa$^{-1}$ and
$K_2=-0.008$ GPa$^{-1}$ and optimal volumes  $V/V_0=0.86$ and
$V/V_0=0.88$, which are in agreement with the ambient pressure
volume deduced from the high pressure experimental data. The
energies in the global minimums in LDA and LDA+U calculations are
270 K and 525 K per TiF$_3$ lower than that of the experimentally
observed structural parameters. A shallower minimum in LDA can be
explained by the kinetic energy loss due to the octahedra tilting.
In LDA+U the kinetic energy is suppressed due to correlations and
does not change much with tilting.

We speculate that this anomalous behavior of the compressibility
of TiF$_3$ may be due to the presence of domain walls in the low
symmetry phase observed in \cite{Mogus}. Under applied pressure
the crystal becomes a single domain. An alternative explanation is
the phonon contribution to the free energy which in principle
depends on pressure. Additional theoretical calculations of the
phonon spectrum and it's pressure depends (Gruneisen parameter) is
needed to estimate the phonon contribution to the bulk modulus
anomaly.

\begin{table}
\caption{\label{tab1} The unit cell parameters for the bulk modulus
calculations.
The normalized volume $V/V_0$ and $c/a$ ratios are taken
from Sowa and Ahsbahs \protect\cite{Sowa}. For LDA+U calculations
we used experimental volume  $V_0=$58.3 \AA$^3$ at ambient pressure
\protect\cite{Sowa} and we choose a smaller
$V_0=$56.2 \AA$^3$ for LDA calculations to match the LDA optimal value.
For given volume and $c/a$ ratios the tilt angle
was chosen such as Ti-F bondlength Eq. (\protect\ref{TiF}) is constant and
equal to $d_{Ti-F}=1.947$ \AA\ \ and $d_{Ti-F}=1.996$ \AA\ \ for LDA and LDA+U
respectively.}
\begin{ruledtabular}
\begin{tabular}{ccccc}
 p (GPa)&$V/V_0$ &c/a &$\phi^0$ LDA& $\phi^0$ LDA+U\\
\hline
0.0001& 1.000 & 1.023 & 10.4 & 13.9 \\
0.46& 0.957 & 1.056 & 14.3 & 17.0 \\
1.05& 0.908 & 1.099 & 17.9 & 20.2 \\
1.62& 0.885 & 1.118 & 19.4 & 21.5 \\
2.37& 0.864 & 1.138 & 20.8 & 22.8 \\
3.18& 0.848 & 1.155 & 21.7 & 23.6 \\
4.03& 0.830 & 1.167 & 22.8 & 24.6 \\
4.97& 0.818 & 1.181 & 23.5 & 25.3 \\
6.05& 0.806 & 1.187 & 24.2 & 25.9 \\
6.40& 0.801 & 1.189 & 24.4 & 26.2 \\
7.23& 0.793 & 1.193 & 24.9 & 26.6 \\
7.67& 0.788 & 1.198 & 25.2 & 26.9 \\
\end{tabular}
\end{ruledtabular}
\end{table}

\section{Conclusion}

We used density functional theory to predict the electronic and
magnetic properties of TiF$_3$. LDA predicts TiF$_3$ to be a
ferromagnetic metal with a fully saturated moment 1 $\mu_B$ per
Ti. The energies of the high and the low symmetry structures are
degenerate. Such that the pure electron-phonon and electrostatic
(Madelung) model does not explain the observed phase transition at
T$_c$=370 K. The correlations are essential to suppress the
kinetic energy loss of Ti $t_{2g}$ electrons to favor the
distortions. To model electron correlations on Ti $d$ orbitals we
use the LDA+U approach, which requires the input parameters $U$
and $J$. We determine these parameters by calculating electron
correlations on the TiF$_6^{2-}$ ion and find $U=8.1$ eV and
$J=0.9$ eV. LDA+U predicts TiF$_3$ to be an antiferromagnetic
insulator with spin $1/2$ per Ti. We find a long range order of Ti
$a_g$ orbitals with a wavevector $(000)$. The low temperature
phase is lower in energy by about 900 K per TiF$_3$ in LDA+U
calculations, which suggests electron-electron correlations are
important.

Using the experimentally determined $c/a$ and $V/V_0$ ratios we
find a global minimum at 14\% (LDA) and 12\% (LDA+U) volumes
smaller than the observed ambient pressure volume, which are
consistent with the ambient pressure volume deduced from the
experimental high pressure data. The experimentally observed
larger volume at the ambient pressure and the unusual behavior of
the bulk modulus could be understood once the phonon contribution
to the free energy is included. The zero phonon energy is in
principle pressure dependent and can alter the position of the
energy global minimum and the bulk modulus. The Gruneisen
parameter calculations and experimental studies of the phonon
spectrum under pressure will help to elucidate light to this
problem.

\begin{acknowledgments}
We are grateful to Fabian Essler for many valuable discussions,
Warren Pickett for help in LDA+U calculations, and Dick Watson
for useful suggestions.
The computations were performed on the BNL galaxy cluster.
This work was supported in part by DOE  Grant No.\ DE-AC-02-98CH10886.
\end{acknowledgments}


\end{document}